\documentclass[12pt]{iopart}
\usepackage{graphicx}
\begin{document}

\title[Polymer solutions]{Polymer solutions: from hard monomers to soft polymers}

\author{Jean-Pierre Hansen, Chris I. Addison and Ard A. Louis}

\address{Dept. Chemistry, Lensfield Rd, Cambridge, CB2 1EW (UK)}
\begin{abstract}
A coarse-graining strategy for dilute and semi-dilute solutions of
interacting polymers, and of colloid polymer mixtures is briefly
described.  Monomer degrees of freedom are traced out to derive an
effective, state dependent pair potential between the polymer centres
of mass.  The cross-over between good and poor solvent conditions is
discussed within a scaling analysis.  The method is extended to block
copolymers represented as ``necklaces'' of soft ``blobs'', and its
success is illustrated here in the case of a symmetric diblock
copolymer which exhibits microphase separation.
\end{abstract}

\maketitle

\section{Introduction: a coarse-graining strategy}

The present paper summarises a collective effort by the Cambridge
group and others to bridge the ``cultural divide'' between statistical
mechanics of polymer solutions and melts on one hand and of ``simple
liquids'' on the other.  While the former is dominated by
field-theoretic methods and scaling concepts pioneered by S.F. Edwards
and P.G. de Gennes, the description of simple liquids, which lack the
scale invariance of polymers, requires a more atomistic approach.  To
bridge the gap we have developed a systematic coarse-graining strategy
first suggested by Flory and Krigbaum\cite{flory}, whereby individual
monomer degrees of freedom are traced out for fixed centre-of-mass
(CM) positions of the polymer coils, thus defining an effective
interaction between the CMs.  Consider a system of N identical
polymer, each consisting of M monomers (or segments) at positions
$\vec r_{i \alpha}$ $(1\leq i \leq N ; 1\leq \alpha \leq M)$.  If
$H(\{\vec r_{i \alpha}\})$ is the interaction hamiltonian of the
system, the probability distribution of the N CMs at positions
$\vec R_i$ is
\begin{eqnarray}
\label{eq1}
P_N(\{\vec R_i\})&
=\frac{1}{Q_N} \int e^{-\beta H(\{\vec r_{i \alpha}\})} \prod_{i=1}^{N}\delta(\vec R_i - \frac{1}{M}\sum_\alpha \vec r_{i \alpha}) \prod_{i=1}^{N} \prod_{\alpha=1}^{M} d \vec r_{i\alpha}\\
&=\frac{e^{-\beta V_{eff}(\{\vec R_i\})}} {\int e^{-\beta V_{eff}(\{\vec R_i\})} \prod_{i=1}^{N} d \vec R_{i}}
\end{eqnarray}
where $\beta=\frac{1}{k_BT}$, $Q_N$ is the partition function for the
$N\times M$ monomers, and the total effective interaction energy of
the N CMs is rigorously defined by:
\begin{eqnarray}
\label{eq2}
V_{eff}(\{\vec R_i\})=-k_BT \ln[C \times P_N(\{\vec R_i\})]
\end{eqnarray}
This effective energy is a free energy, and hence state-dependent,
and, in general, many-body in nature.  In the low concentration limit,
(\ref{eq2}) reduces to a sum of effective pair interactions between
two isolated polymer coils:
\begin{eqnarray}
\label{eq3}
v_2(|\vec R_1 -\vec R_2|)=-k_BT \ln[C \times P_2 (\vec R_1,\vec R_2)]
\end{eqnarray}
Since global properties of polymeric systems are independent of
chemical detail in the scaling ($L\rightarrow \infty$) limit, we adopt
henceforth a simple lattice model of polymer solutions namely that of
mutually and self-avoiding walks (SAW) of length $L=M-1$ on a cubic
lattice of lattice spacing $b$ (equal to the segment length);
non-connected nearest neighbour monomers of the same or different
polymer coils have an attractive energy $-\epsilon$.  This model
accounts for the key polymer features, namely connectivity, excluded
volume and solvent quality (through the value of $\epsilon$).

Convenient thermodynamic variables are the polymer density
$\rho=N/(\Omega b^3)$ ($\Omega$ being the size of the lattice), the
monomer density $c=M\rho$ (equivalently the monomer packing fraction
$\phi=cb^3$ equal to the number of lattice sites occupied by monomers)
and the temperature T ($\beta^*=\epsilon/k_BT)$.  A key
characteristic is the overlap density $\rho^*=3/4\pi R{_g}{^3}$
(where $R_g$ is the radius of gyration, $\sim bL^\nu$) which corresponds to the
cross-over from the dilute ($\rho <\rho^*$) to the semi-dilute regimes
($\rho >\rho^*$).  The semi-dilute regime differs from the melt in
that the monomer packing fraction remains negligible; for any given
$\rho/\rho^*$ this is only achieved for sufficiently long polymers,
since $\phi=(\rho/\rho^*) L^{(1-3\nu)} \sim L^{-4/5}$ for SAW polymers.

\section{Polymers in good solvent}
Consider first the athermal limit of SAW polymers ($\epsilon=0$).  In
the low ($\rho\rightarrow 0$) density limit $P_2 (\vec
R_1,\vec R_2)$ is then simply the probability that there is
no monomer-monomer overlap for a fixed distance $r=|\vec
R_1 -\vec R_2|$ between the CMs.  This is well adapted to
Monte-Carlo (MC) sampling.  Scaling theory predicts that the resulting
effective interaction at full overlap, $v_2(r=0)$ is independent of
chain length L\cite{gros}.  Early simulations with rather short
chains pointed to $v_2(r=0) \approx 2k_BT$\cite{daut}.  An in-depth
investigation of the L-dependence shows that $v_2(r)$ is well
represented by a Gaussian $v_2(r) \approx u
exp(-\alpha(r/R_g)^2)$, where $\alpha$ is of order 1, while
$u/k_BT \approx 1.80$ \cite{bolh}\cite{pelis}.\\ At finite
polymer concentration $\rho$, three and more-body effective
interactions come into play\cite{bolh2}. A more efficient strategy is
to determine an effective density-dependent pair potential
$v_2(r;\rho)$ by inverting the CM-CM pair distribution $g(r)$ from full
monomer-level MC simulations\cite{bolh2}\cite{louis}.  It was proven
that this inverse problem has a unique solution\cite{hend}.  The
inversion procedure is implemented using the HNC-integral
equation\cite{hansen}. The Ornstein-Zernike relation\cite{hansen}
allows the extraction of the direct correlation function $c(r)$ from
the MC data for $h(r)=g(r)-1$ at any given density.  The HNC closure then expresses $v_2(r)$ as \cite{hansen}:
\begin{eqnarray}
\label{eq4}
\beta v_2(r)=-\ln[g(r)]+h(r)-c(r).
\end{eqnarray}
The first term on the r.h.s. is the potential of mean force, while
$h(r)-c(r)$ describes the effect of correlations.\\ 

The MC generated $g(r)$ show that correlations {\it decrease} as $\rho$
increases, contrary to the more familiar behaviour observed for hard
core systems.  The overlap value $g(r=0)$ increases steadily toward 1
(the ideal gas value) confirming that in the high density limit of a
melt, polymer chains indeed behave as non-interacting
polymers\cite{rubin}.  The resulting effective pair potential is only
moderately density dependent and is well fitted by a sum of gaussians.
The range of $v_2(r)$ tends to increase with $\rho$, and the potential
develops a small amplitude negative tail for $r$ significantly larger
than $R_g$\cite{bolh}\cite{bolh3}.\\ 
The link with thermodynamics is via the
compressibility relation\cite{hansen}, which allows the osmotic
pressure $P$ to be expressed as:
\begin{eqnarray}
\label{eq5}
\beta P(\rho)= \int_{0}^{\rho}[1-\rho'\hat{c}(k=0;\rho')]d\rho'  
\end{eqnarray}
where $\hat{c}(k)$ is the Fourier transform of $c(r)$.  Use of the
effective pair potential in conjunction with the virial and energy
equations is meaningless\cite{louis5}.  The equations of state
calculated from MC simulations of the full monomer level polymer
representation and from simulations based on the much less
CPU-intensive effective potential representation agree within
numerical uncertainties, underlining the adequacy if the HNC inversion
procedure for such ``soft'' effective particles.  Well into the
semi-dilute regime ($\rho \gg \rho^*$) the slopes of the calculated
equation of state agrees with the des Cloizeaux scaling prediction
$\beta P \approx \rho^{3\nu/(3\nu -1)} \approx \rho^{9/4}$, where
$\nu=0.588 \approx 3/5$ is the Flory exponent for the radius of
gyration in good solvent($R_g \sim bL^\nu)$\cite{rubin}.\\
Neglecting the density-dependence of $v_2(r)$, {\it i.e.} extending
the low density gaussian form to all densities, brings us back to the
``Gaussian core model'' (GCM) first introduced by
Stillinger\cite{stil}, which exhibits interesting behaviour at low
temperatures ($\beta^* \gg 1)$\cite{lang}.  In the regime relevant for
polymer solutions ($\beta^* \approx 1)$ the model leads to ``mean
field fluid'' behaviour at sufficiently high density, where the random
phase approximation, $c(r)=-\beta v_2(r)$ leads to a quadratic
equation-of-state for $\rho \gg \rho^*$\cite{lang}\cite{louis2}:
\begin{eqnarray}
\label{eq6}
\beta P=\rho+ \frac{1}{2}\beta \hat{v}_2(k=0)\rho^2
\end{eqnarray}
Incorporating the $\rho$-dependence of $v_2$ is thus seen to change
the asymptotic $\rho^2$ behaviour into des Cloizeaux scaling
$\rho^{9/4}$.  As a by-product of the GCM, we have developed a
multiple occupancy lattice model, which also gives rise to interesting
microphase separation\cite{finken}.

\section{From good to poor solvent conditions}
We now turn our attention to the case where adjacent monomers attract,
{\it i.e.} $\epsilon\neq 0$ or $\beta^*>0$.  This
attraction is solvent induced, and the quality of the solvent
deteriorates as $\beta^*$ increases, leading to contraction of the
polymer coils.  At the $\theta$ temperature ($\beta^*_\theta =
\epsilon/k_BT_\theta$), repulsion and attraction between polymers
cancel, at least in the low density limit, so that polymers exhibit
the scaling behaviour of ideal polymers($R_g \sim L^{1/2}$).  Below
$T_\theta$, polymer coils collapse into globules ($R_g \sim L^{1/3}$),
and phase separation occurs into polymer-rich and polymer-poor
solutions. \\ 
The most convenient diagnostic for locating $T_\theta$
from simulations is to calculate the second virial coefficient
$B_2(L;T)$ as a function of temperature
and polymer length.  The L-dependent Boyle temperature $T_B(L)$ is
that at which $B_2(L;T)$ vanishes for a fixed L, then:
\begin{eqnarray}
\label{eq7}
T_\theta=\lim_{L \rightarrow \infty}T_B(L)
\end{eqnarray}
This leads to the estimate $\beta^*_\theta=0.2690 \pm 0.0002$
\cite{grass}.  Note that $B_2(L;T)$ can be directly expressed in
terms of the low density limit of the effective CM pair potential:
\begin{eqnarray}
\label{eq8}
B_2(L;T)=2\pi \int_0^\infty[1-e^{-\beta v_2(r, L, T)}]r^2dr
\end{eqnarray}
Extensive MC simulations were used to determine the effective pair
potential for fixed length $L=100$ over a wide range of
temperatures($0\leq\beta^* \leq 0.3$) and densities $\rho$\cite{krak}.
As $\beta^*$ increases, $v_2(r=0)$ decreases and $v_2(r)$ develops an
attractive tail for $r > R_g$.  Eventually $v_2(r)$ violates Ruelle's
necessary condition for the existence of a thermodynamic limit,
namely\cite{ruel}:
\begin{eqnarray}
\label{eq9}
I_2=\int v_2(r) d \vec{r} >0 
\end{eqnarray}
Since $v_2$ depends on $L$, $\rho$ and $T$, so does $I_2$, and for any given $\rho$ and $L$, the limit of stability temperature $T_s$ is determined by 
\begin{eqnarray}
\label{eq10}
I_2(L,\rho, T=T_s)=0
\end{eqnarray}
Below $T_s$, single polymer coils will collapse and we conjecture that
$T_\theta=\lim_{\rho \rightarrow 0} \lim_{L \rightarrow \infty}
T_s(\rho, L) $\cite{krak}.  However, if $\rho$ is increased at a given
temperature $T\leq T_\theta$ the effective pair potential increasingly
reverts to being repulsive until the Ruelle criterion\cite{hansen} is
satisfied, and the polymer solution becomes thermodynamically stable
again. This ``restabilisation'' reflects a phase separation scenario
under poor solvent conditions($T < T_\theta$). \\ The variation of
$v_2(r)$ with temperature and density, as extracted from the HNC
inversion of MC data, is semi-quantitatively reproduced by solutions
of the PRISM integral equation\cite{schw} for the thread
model($b\rightarrow0, L\rightarrow\infty$ at fixed $R_g$) of polymer
solutions\cite{fuchs}\cite{krak2}.  PRISM assumes all monomers to be
equivalent ({\it i.e.} neglects ends effects) and yields the monomer
pair distribution function $g_{mm}(r)$ of a polymer solution.  The CM
distribution function $g(r)$ required to extract the effective pair
potential $v_2(r)$ may be related to $g_{mm}(r)$ by an approximate,
but accurate relation involving internal form factors of a single
coil\cite{krak3}.  Further progress along these lines might eventually
bypass the need for time-consuming simulations of full monomer level
models required to determine $g(r)$.  \\ The equation of state may be
computed as a function of $\rho/\rho^*$ and $\beta^*$, using a
somewhat cumbersome method based on the contact
theorem\cite{dick}\cite{add}, or much more efficiently by subjecting
the polymers to a gravitational field and invoking hydrostatic
equilibrium\cite{add2}.  If $\rho(z)$ denotes the CM(or monomer)
density profile of the polymers in a vertical field, which is easily
measured in simulations, the osmotic pressure at an altitude $z$ is
simply
\begin{eqnarray}
\label{eq11}
\beta P(z)= \frac{1}{\zeta}\int_z^\infty \rho(z')dz'
\end{eqnarray}
where $\zeta=k_BT/Mg$ is the gravitational length, which must be
chosen (by tuning the product $Mg$) significantly larger than $R_g$
for the macroscopic description (which follows from the local density
approximation within density functional theory of inhomogeneous
fluids) to hold\cite{barrat}.  Elimination of the altitude $z$ between
$P(z)$ and $\rho(z)$ then leads to the bulk equation of state
$P(\rho)$. Examples from simulations of L=500 chains are shown, for
four temperatures in figure 1.
\begin{figure}[!htp]
\begin{center}
\includegraphics[width=8cm]{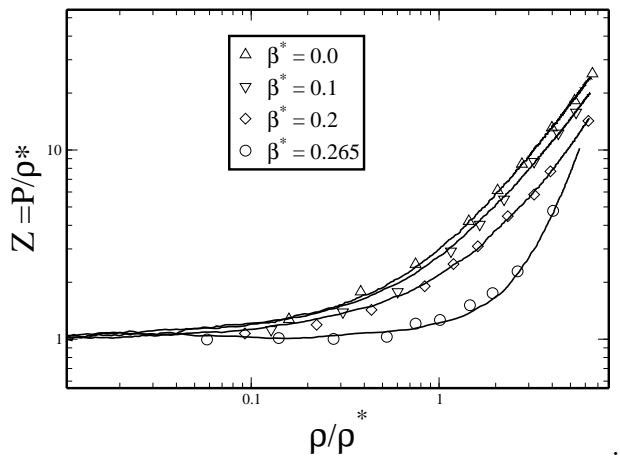}.
\caption{Equation of state on a log-log scale, calculated using Dickman's(contact theorem) method, (symbols) and hydrostatic equilibrium(solid lines).}
\label{fig1}
\end{center}
\end{figure}
\section{Long polymers: corrections to scaling}
Polymers are critical objects in the universality class of the zero
component(n=0) limit of the n-vector model of critical
phenomena\cite{genne}.  The $L\rightarrow \infty$ limit is equivalent
to the limit of divergent correlation length $\xi$ at the critical
point of a second-order phase transition.  The properties of very long
chains can hence be investigated using the powerful method of the
renormalization group (RG) and field theory\cite{freed}.  These
predict two scaling regimes, an athermal one corresponding to the good
solvent (SAW) limit, and the second corresponding to the
$\theta$-solvent regime where ideal polymer statistics hold.  The
cross-over between the two regimes is discontinuous in the scaling
limit $L \rightarrow \infty$.  The objective is to predict the
behaviour for large but finite L from a finite size scaling analysis
of MC data\cite{li}\cite{pelis}.  We restrict the discussion to the
good solvent regime.  For $\beta<\beta_\theta$ any universal
(dimensionless) ratio $R$ may be represented as
\begin{eqnarray}
\label{eq12}
R(L,\beta)=R^* +\frac{a_R(\beta)}{L^\Delta} + ...
\end{eqnarray}
where $R^*$ is the temperature independent scaling limit
($L\rightarrow \infty$) and the exponent $\Delta=0.517$\cite{beloh}.
Higher order terms involve exponents of the order of 1 or larger.  The
temperature-dependent coefficient $a_R(\beta)$ is non-universal, but
ratios of such coefficients for two different dimensionless qualities
are again universal, {\it i.e.} model-independent.  A much studied
example of a universal ratio is
$A_2(L,\beta)=B_2(L,\beta)/R^3_g(L,\beta)$.  A similar ratio involving
the third virial coefficient
$A_3(L,\beta)=B_3(L,\beta)/R^6_g(L,\beta)$ has been examined in detail
in reference\cite{pelis}.  $B_3$ is found to be always positive, but
to go through a sharp minimum near $T=T_\theta$.  The effective pair
potential between the CMs of two isolated polymers ($\rho \ll \rho^*$
limit), divided by $k_BT$, is another universal ratio, which has been investigated by a similar scaling analysis in \cite{pelis}:
\begin{eqnarray}
\label{eq13}
\beta v_2(r,L,\beta)=v_\infty(x) +\frac{a_v(\beta)}{L^\Delta}v_c(x)  + ...
\end{eqnarray}
where $x=r/R_g$.  $v_\infty$, $v_c$ and $a_v(\beta)$ are extracted
from a careful analysis of MC data for several lengths L and inverse
temperatures $0 \leq \beta < \beta_\theta$.  These may then be used to
predict $\beta v_2$ for any length and temperature, and agreement with
MC data is excellent for $\beta^* \leq 0.2$ and $L\geq 500$.  Closer
to the $\theta$-temperature the convergence of (\ref{eq13}) with L is
found, not surprisingly, to be much slower, and one must then switch
to the scaling analysis appropriate for the $\theta$-regime.  The
scaling analysis has been recently extended to binary mixtures of
polymers of different lengths and to star polymers\cite{pelis3}

\section{Soft polymers and hard colloids}
Mixtures of colloidal particles and non-adsorbing polymers have
attracted considerable experimental and theoretical attention over the
last two decades, because of interesting phase behaviour induced by
the familiar depletion mechanism\cite{asaku}.  The effective depletion
interaction between spherical colloidal particles may be tuned by
varying the size ration $q=R_g/R_c$ (where $R_c$ is the colloid
radius), and the polymer concentration $\rho$.  Consider first
the case of polymers between 2 plates($q=0$) separated by z.  The
depletion potential per unit area $W(z)$ is defined by the difference
in polymer grand potentials:
\begin{eqnarray}
\label{eq14}
W(z)&=\frac{1}{A}[\Omega(z)-\Omega(z=\infty)]\\
&=\int_z^\infty[P(z')-P(z=\infty)]dz'
\end{eqnarray}
Clearly since at contact 2 depletion zones are destroyed
$W(z=\infty)=-2\gamma_W(\rho)$, where $\gamma_W$ is the polymer -wall
surface tension.  The simple ansatz: 
\begin{eqnarray}
\label{eq15}
W(z)&=W(0) + P(\rho)z  &\;\;\;\;\;\; ; z<D_w(\rho)=-\frac{W(0)}{P(\rho)}\\
&    =0 &               \;\;\;\;\;\; ; z>D_w(\rho)
\end{eqnarray}
reproduces direct simulation data well.  The result for SAW
polymers\cite{louis3} differs considerably from that for ideal
polymers\cite{asaku}: the range $D_W$ is shorter for interacting
polymers, and decreases with increasing density, while it is
independent of density for ideal polymers; the contact value $W(0)$
decreases faster with $\rho$ for interacting polymers since the
surface tension scales as $\rho^{3/2}$ in the semi-dilute
regime\cite{louis4} while it scales as $\rho$ for ideal polymers.  For
finite $q\leq 1$, the depletion force between two spheres can be
approximately related to the depletion potential between 2 planes via
the Derjaguin approximation, and by correcting for the decreasing
range of the force due to partial wrapping of the polymer coils around
the spherical colloid\cite{louis3}.\\ Such pair depletion interactions
do not, however, account for effective many-body interactions due to
finite colloid concentrations.  To that purpose the coarse-graining
strategy described in sections 1 to 3 may be extended to the two
component colloid-polymer system.  An effective state-dependent
colloid-polymer pair interaction $v_{cp}(r)$ may be extracted by an
HNC inversion of the polymer density profile around a
sphere\cite{bolh}\cite{bolh3}\cite{louis3}, similar to the inversion
procedure used to determine the effective polymer-polymer pair
potential $v_{pp}(r)$.  The colloid-colloid pair potential $v_{cc}(r)$
is well approximated by a simple hard sphere interaction.  MC
simulations of this effective two-component system were used to
calculate the phase diagram of the mixture for $q \leq 1$
\cite{bolh4}.  The calculated binodal agrees very well with
experimental data\cite{rama} for interacting polymers. Significant
qualitative differences arise between phase diagrams for
ideal\cite{lekker} and interacting polymers, particularly for larger
$q$:  the range of the concentrated (``liquid'') colloidal phase in
the colloid density-polymer density plane is considerably reduced when
polymer interactions are included, and the critical point occurs at
significantly higher packing fraction of the two species\cite{bolh4}.
These trends become more pronounced in the ``protein limit'' of large
polymers and small colloids ($q\gg 1$)\cite{bolh5}.  Solvent quality
has a strong influence on the induced depletion interaction between
colloids, with polymers under $\theta$ conditions leading, not
surprisingly to a pair interaction close (but not identical) to that
induced by ideal polymers\cite{add}, at least for $\rho/\rho*<1$.

\section{Diblock copolymers and beyond}
The coarse-graining strategy may be extended to polymers other than
linear homopolymers considered so far.  Star polymers, for instance,
have been investigated along similar lines\cite{likos}; the CM is
replaced by the midpoint where the $f$ arms of a star polymer meet;
the resulting effective pair interaction diverges logarithmically as
$r\rightarrow0$ and hardens as $f$ increases. \\ The case of symmetric
diblock copolymers AB has been examined very recently\cite{add3}.  The
A and B strands are represented as soft ``blobs'', the CMs of which
are tethered by an ``entropic spring'' characterised by the
intramolecular potential $\phi_{AB}(r)$ which is derived from the
MC-generated distribution function of relative distances of the A and
B CMs on the same copolymer.  There are now three intermolecular CM
potentials $v_{AA}(r)$, $v_{BB}(r)$ $v_{AB}(r)$, in addition to the
intramolecular potential.  The inversion procedure to go from the
partial distribution functions $g_{\alpha \beta}(r)$ to the pair
potentials $v_{\alpha \beta}(r)$ is now more involved\cite{add3}.
Even in the low density limit one already faces a four-body problem.
The inversion procedure has been carried out in that limit for a
simple athermal model, the ISI model, where A-A and B-B pairs behave
like ideal polymers, {\it i.e.} freely interpenetrate, while A-B pairs
behave like mutually avoiding walks. This is the block copolymer
equivalent of the familiar Widom-Rowlinson model\cite{widom} which
drives phase separation of simple atomic fluids (where A and B are
untethered).  Because the two strands of the AB block copolymer system
are tethered, macroscopic phase separation due to A-B incompatibility
is suppressed and reduces to microphase separation: the symmetric
block copolymers form a lammelar phase, which was indeed observed in
MC simulations of both the full monomeric and the coarse-grained
representations\cite{add3}.  The resulting equation of state calculated
by the hydrostatic equilibrium method is shown in Figure 2.
\begin{figure}[!htp]
\begin{center}
\includegraphics[width=8cm,angle=0]{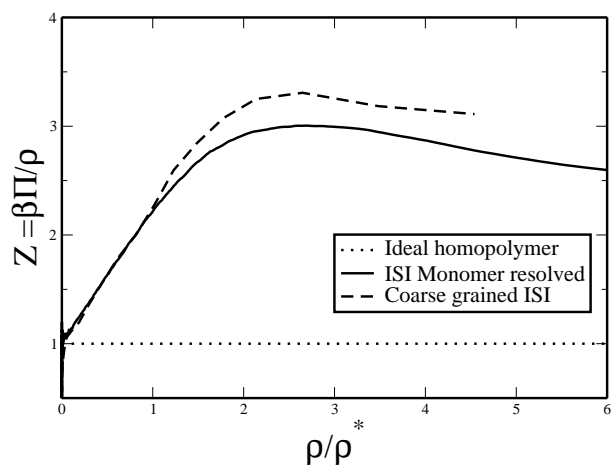}
\caption{Equation of state generated from MC simulations for diblock copolymers
using full monomer level and coarse-grained ``blob'' models.}
\label{fig2}
\end{center}
\end{figure}
$Z=\beta P/\rho$ is seen to first increase linearly up to
$\rho/\rho^*\approx2$, where it flattens out, and thereafter decreases
slowly to give an asymptotic value $>1$.  This may be understood by
noting that in the lamellar phase(which develops for $\rho/\rho^*>2$),
the repulsive A-B contacts are greatly reduced.  Figure 2 also shows
the equation of state calculated (with much less computational effort)
for the coarse-grained ``soft-dumbbell'' model with effective intra
and inter-molecular pair potentials $\phi_{AB}(r)$ and $v_{\alpha
\beta}(r)$ determined in the zero density limit.  The agreement is
excellent up to $\rho\approx\rho^*$, and remains semi-quantitative
thereafter, despite the fact that the density dependence of the
effective potential has not been taken into account.

The ``soft dumbbell'' representation of diblock copolymers provides a
hint of how to extend the coarse-graining strategy of linear homo or
hetropolymers over a wide range of polymer concentrations.  As the
ratio $\rho/\rho^*$ increases, the fundamental length scale gradually
crosses over from $R_g$ (for $\rho/\rho^*<1$) to the correlation
length $\xi \sim \rho^{-3/4}$ deep in the semi-dilute regime, to the
segment length $b$ in the melt.  The coarse-graining strategy put
forward in this paper applies to dilute and initial semi-dilute
regime, where polymer coils are well represented by a single
``soft-core'' particle with a radius of the order $R_g$.  Deeper into
the semi-dilute regime, the blob picture\cite{genne2} applies for
polymers confined by other polymers or in a pore; each polymer reduces
to a ``necklace'' of blobs of radius $\xi$, tethered by entropic
springs; different blobs on the same or neighbouring chains interact
via a quasi-gaussian soft-core potential, as introduced in earlier
sections.  The radius of each blob decreases, and hence their number
increases (for a given overall length L) as the ratio $\rho/\rho^*$
increases, until the melt regime is reached where the blob size
reduces essentially to the segment length $b$, and the coarse-graining
strategy is no longer of any use. Over the whole semi-dilute regime a
polymer may thus be pictured as a necklace of blobs, and the present
coarse-graining strategy allows, in principle for an unequivocal
determination of intra and inter-molecular effective interactions
between these blobs.  Work along these lines is in progress.

\section{Acknowledgements}
The authors are grateful to their collaborators on this project over
the years; the work presented in this overview owes much to Peter
Bolhuis, Andrea Pelissetto, Vincent Krakoviack, Reimar Finken,
Evert-Jan Meijer and Benjamin Rotenberg.  CIA acknowledges the support
of the EPSRC and AAL is grateful to the Royal Society of London for their support.


\section{References}
\begin{thebibliography}{99}
\bibitem{flory} P.J. Flory and W.R. Krigbaum, {\it J.Chem.Phys.} {\bf 18}, 1086 (1950)
\bibitem{gros} A.V. Grosberg, P.G. Khalatur and A.R. Khokhlov, {\it Macromol.Chem.RapidCommun.} {\bf 3}, 709 (1982)
\bibitem{daut} J. Dautenhahn and C.K. Hall, {\it Macromolecules} {\bf 27}, 5399 (1994)
\bibitem{bolh} P.G. Bolhuis, A.A. Louis, J.-P. Hansen and E.J. Meijer {\it J.Chem.Phys.} {\bf 114}, 4296 (2001)
\bibitem{pelis} A. Pelissetto and J.-P. Hansen, {\it J.Chem.Phys.} {\bf 112}, 134904 (2005)
\bibitem{bolh2} P.G. Bolhuis, A.A. Louis and J.-P. Hansen, {\it Phys.Rev.E} {\bf 64}, 021801 (2001)
\bibitem{louis} A.A. Louis, P.G. Bolhuis, J.-P. Hansen and E.J. Meijer, {\it Phys.Rev.Letters} {\bf 85}, 2522 (2000)
\bibitem{hend} R.L. Henderson, {\it Phys.Lett.A} {\bf 49}, 197 (1974)
\bibitem{hansen} J.-P. Hansen and I.R. McDonald, ``Theory of Simple Liquids'' $2^{nd}$ ed, (Academic Press, London, 1986)
\bibitem{rubin} see {\it e.g.} M. Rubinstein and R.H. Colby, ``Polymer Physics'', (Oxford University Press, 2003)
\bibitem{bolh3} P.G. Bolhuis and A.A. Louis, {\it Macromolecules} {\bf 35}, 1860 (2002)
\bibitem{louis5} A.A. Louis,  {\it J. Phys.Cond.Matt} {\bf 14} 9187 (2002)  
\bibitem{stil} F.H. Stillinger, {\it J.Chem.Phys.} {\bf 65}, 3968  (1976)
\bibitem{lang} A. Lang, C.N. Likos, M Watzlawek and H. L\"owen, {\it J.Phys.Cond.Matt} {\bf 12}, 5087 (2000)
\bibitem{louis2} A.A. Louis, P.G. Bolhuis and J.-P. Hansen, {\it Phys.Rev.E} {\bf 62}, 7961 (2000)
\bibitem{finken} R. Finken, J.-P. Hansen and A.A. Louis, {\it J.Phys.A} {\bf 37}, 577 (2004)
\bibitem{grass} P. Grassberger and R. Hegger, {\it J. Chem. Phys.} {\bf 102}, 6881 (1995)
\bibitem{krak} V. Krakoviack, J.-P. Hansen, and A.A. Louis, {\it Phys.Rev.E} {\bf 67}, 041801 (2003)
\bibitem{ruel} D. Ruelle ``Statistical Mechanics: Rigorous Results'', (Benjamin, London, 1969)
\bibitem{schw} For a review see K.S. Schweizer and J.G. Curro, {\it Adv.Chem.Phys.} {\bf 97}, 1 (1997)
\bibitem{fuchs} M. Fuchs and K.S. Schweizer, {\it Phys.Rev.E} {\bf 64}, 021514  (2001)
\bibitem{krak2} V. Krakoviack, B. Rotenburg and J.-P. Hansen {\it J.Phys.Chem.B} {\bf 108}, 6697 (2004)
\bibitem{krak3} V. Krakoviack, J.-P. Hansen, and A.A. Louis, {\it Europhys.Lett.} {\bf 58}, 53 (2002)
\bibitem{dick} R. Dickman, {\it J. Chem. Phys.} {\bf 87}, 2246 (1987)
\bibitem{add} C.I. Addison, A.A. Louis and J.-P. Hansen, {\it J.Chem.Phys.} {\bf 121}, 612 (2004)
\bibitem{add2} C.I. Addison, J.-P. Hansen and A.A. Louis, {\it Chem.Phys.Chem.} (in press 2005)
\bibitem{barrat} J.-L. Barrat, T. Biben and J.-P. Hansen, {\it J.Chem.Phys.} {\bf 102}, 6881 (1995)
\bibitem{genne} P.G. de Gennes, {\it Phys.Lett} {\bf 38A}, 339 (1972)
\bibitem{freed} see {\it e.g.} K.F. Freed, ``Renormalization Group Theory of Macromolecules'',  (Wiley, New York, 1987)
\bibitem{li} B. Li, N. Madras and A.D. Sokal, {\it J.Stat.Phys.} {\bf 80}, 661 (1995)
\bibitem{beloh} P. Belohorec and B. Nickel, Guelph University Report (1997) (unpublished)
\bibitem{pelis3} A. Pelissetto, {\it Private communication}
\bibitem{asaku} S. Asakura and F. Oosawa, {\it J.Chem.Phys.} {\bf 22}, 1255 (1954)
\bibitem{louis3} A.A. Louis, P.G. Bolhuis, E.J. Meijer and J.-P. Hansen, {\it J.Chem.Phys.} {\bf 117}, 1893 (2002)
\bibitem{louis4} A.A. Louis, P.G. Bolhuis, E.J. Meijer and J.-P. Hansen, {\it J.Chem.Phys.} {\bf 116}, 10547 (2002)
\bibitem{bolh4} P.G. Bolhuis, A.A. Louis and J.-P. Hansen, {\it Phys.Rev.Lett} {\bf 89}, 128302 (2002) 
\bibitem{lekker} H.N.W. Lekkerkerker {\it et al.}, {\it Europhys.Lett} {\bf 20}, 559 (1992); E.J. Meijer and D. Frenkel, {\it J.Chem.Phys.} {\bf 100}, 6873 (1994)
\bibitem{rama} S. Ramakrishman, M. Fuchs, K.S. Schweizer and C.F. Zukoski, {\it J.Chem.Phys.} {\bf 116}, 2201 (2002)
\bibitem{bolh5} P.G. Bolhuis, E.J. Meijer and A.A. Louis, {\it Phys.Rev.Lett} {\bf 90}, 068304 (2003)
\bibitem{likos} For a review, see C.N. Likos, {\it Phys.Rep} {\bf 348}, 267 (2001)
\bibitem{add3} C.I. Addison, J.-P. Hansen, V. Krakoviack and A.A. Louis, {\it Mol.Phys.} (in press 2005)
\bibitem{widom} B. Widom and J.S. Rowlinson, {\it J.Chem.Phys.} {\bf 52}, 1670 (1970)
\bibitem{genne2} P.G. de Gennes, "Scaling Concepts in Polymer Physics" (Cornell University Press, Ithaca 1979)

\end{thebibliography}
\end{document}